\documentstyle[amssymb,twocolumn,prl,aps,graphicx]{revtex}
\begin{document}
\twocolumn[ \hsize\textwidth\columnwidth\hsize\csname
@twocolumnfalse\endcsname

\title{
       Evidence of two viscous relaxation processes \\ 
	in the collective dynamics of liquid lithium.
      }
\author{
        T.~Scopigno$^{1}$,
        U.~Balucani$^{2}$,
        G.~Ruocco$^{3}$,
        F.~Sette$^{4}$
        }
\address{
    $^{1}$Dipartimento di Fisica and INFM, Universit\'a di Trento, I-38100, Povo, Italy.\\
    $^{2}$Istituto di Elettronica Quantistica CNR, I-50127, Firenze, Italy. \\
    $^{3}$Dipartimento di Fisica and INFM, Universit\'a di L'Aquila, I-67100,L'Aquila, Italy.\\
    $^{4}$European Synchrotron Radiation Facility, B.P. 220 F-38043 Grenoble, Cedex France.
    }
\date{\today}
\maketitle

\begin{abstract}
New inelastic X-ray scattering experiments have been performed on
liquid lithium in a wide wavevector range.
With respect to the previous measurements, the instrumental
resolution, improved up to 1.5 meV, allows to accurately
investigate the dynamical processes determining the observed
shape of the the dynamic structure factor, $S(Q,\omega)$.
A detailed analysis of the lineshapes shows
the co-existence of relaxation processes with both a slow and a fast
characteristic timescales, and therefore that pictures
of the relaxation mechanisms based on a simple viscoelastic model
must be abandoned.
\end{abstract}


]

The collective dynamics of liquid alkali metals
exhibits several features which make these systems excellent candidates to
test different theories on the microdynamics of the liquid
state. In simple liquids, both generalized kinetic theory and
mode-coupling treatments predict the existence of two distinct relaxation
processes affecting the dynamics of density fluctuations \cite{bal}. In
these theories the dominant damping mechanism is provided by a fast process
which is thought to be associated with the interactions between an atom and 
the ''cage'' of its nearest neighbors. In addition, one should also
detect a much slower process involving the cooperative motion of a
larger number of particles, which rearranges the local structure to
ultimately restore equilibrium, and which is responsible
for the critical slowing down in those systems capable of supercooling.
In simple liquids such as the molten alkali metals most of these
predictions have been tested by molecular dynamics (MD) simulations\cite{ram}.
In these numerical studies several phenomena have been detected, such
as the persistence of well-defined density modes outside the strict
hydrodynamic region and an increase of the sound velocity in the mesoscopic
wavevector region (the so-called ''positive dispersion''). More recently,
considerable attention has been devoted to check the existence of the 
afore-mentioned relaxation mechanisms \cite{now,can}.

Up to few years ago, the only experimental technique capable to
access the relevant wavevector/frequency range was inelastic
neutron scattering (INS). Studies were reported on rubidium
\cite {cop}, cesium \cite{bod}, lithium
\cite{ver,tor}, potassium \cite{nov} and again rubidium
\cite{chi}. Probing wavevectors $Q$ up to the first sharp diffraction
peak, collective excitations outside the strict hydrodynamic
regime were detected. The INS data have usually been
analyzed with very simple models and, in the favorable cases 
(scattering cross-section mostly coherent and speed
of sound low enough to overcome kinematics restrictions), 
it was possible to observe the positive dispersion 
of the sound velocity of the acoustic mode - one of the features predicted by the MD
simulations. This situation is exemplified by the INS experiments
of Bodensteiner et al. in liquid cesium \cite{bod}, where the
quality of the data allowed even to test the lineshape predicted
by the viscoelastic model, a sort of simplified relaxation
mechanism with a single (average) decay time. However, even in
this case, the data accuracy did not allow to test whether the 
two (physically quite different) mechanisms mentioned previously 
appear in the experimental determination of the $S(Q,\omega)$.

The development of inelastic X-ray scattering (IXS)
technique, paved the way for the experimental investigations of
the collective dynamics of liquids in the mesoscopic  region. 
The aim of the present work is to report very high quality IXS 
data of the $S(Q,\omega )$ of liquid lithium, a case particularly 
unfavourable to INS but showing remarkable inelastic features \cite{sinn}. 
These data allow to accurately study the spectral shape, and to 
obtain the first strong evidence on the existence of a double-timescale decay
mechanism underlying the dynamics of density fluctuations. The
adoption of a spectral analysis scheme based on the generalized
Langevin equation formalism, enables us to extract quantitative
information on the relaxation processes.

The IXS experiment was carried out at the high resolution beamline ID16 of the
European Synchrotron Radiation Facility (Grenoble, F). The back- scattering 
monochromator and analyser crystals, operating at the $(h h h)$ silicon
reflections with $h$$=$$9,11$, gave a total energy resolution of 3 meV 
for $h$$=$$9$ and 1.5 meV for $h$$=$$11$.
The wavevector transferred in the scattering process was selected 
between 1.4 nm$^{-1}$ and 25 nm$^{-1}$ by rotating a 7 m
long analyzer arm in the horizontal scattering plane.
The total $Q$ resolution was set to 0.4 nm$^{-1}$.
Energy scans, performed by varying the temperature of
the monochromator with respect to that of the analyzer crystals, took 
about 180 min, and each spectrum at a given $Q$ was obtained from the 
average of 2 to 8 scans, depending on the values of $h$ and of $Q$.
The sample (Goodfellows) had a nominal purity better than 0.001.
The liquid lithium uncapped container was made of austenitic stainless
steel, and a resistance heater was used to keep the
liquid at 475 K, i.e. slightly above the melting point at 453 K.
Additional data were collected at 600 K.
The 20 mm long sample, kept together by
surface tension, was maintained in a 10$^{-6}$ mbar vacuum
and loaded in an argon glove box.
In the $Q-E$ region of interest, empty vacuum chamber
measurements gave either the flat electronic detector background of 0.6
counts/min ($Q$$>$8 nm$^{-1}$), or, for $Q$$<$8 nm$^{-1}$ a small elastic
line due to scattering from the chamber kapton
windows (each 50 $\mu$m thick). This signal, after proper normalization,
has been subtracted from the raw data. The scattered intensity $I(Q,\omega )$ was reduced to absolute
units following a procedure described in Ref.\cite{scop}, and the 
contribution from multiple scattering was negligible \cite{multnote}.

Typical spectra are reported in Fig.~1 (full dots).

\vspace{-.46cm}
\begin{figure}[ftb]
\hspace{-.75cm}
\centering
\includegraphics[width=.43\textwidth]{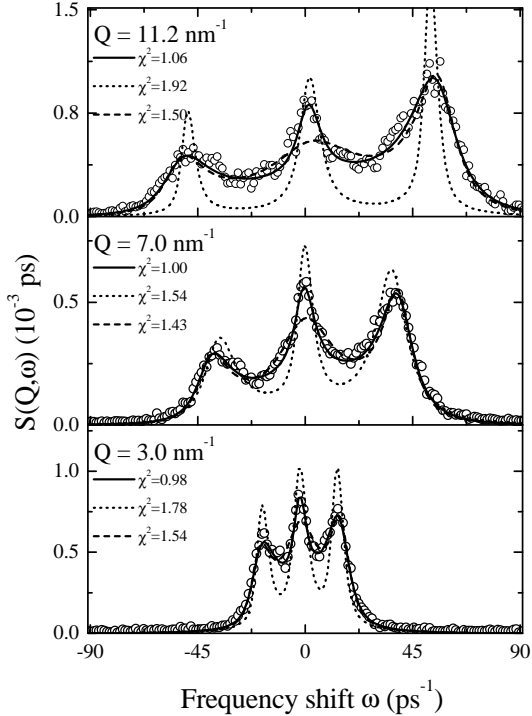}
%\vspace{-.05cm}
\caption{
IXS spectra of liquid lithium at $T=475$
K at three wavevectors. Open Dots: experimental data. Dashed/Dotted lines:
results of two fitting sessions (resolution convoluted) with comparable
$\chi ^2$ for a memory function model which includes thermal decay and a
single-time viscous relaxation channel (viscoelastic model). Full line:
same as before, except that now the viscous relaxation proceeds with two
decay channels (microscopic and structural relaxation). Normalized $\chi^2$ 
values are also reported, the expected value is $1 \pm 0.1$.}
\label{Fig1}
\end{figure}

The data analysis has been performed starting from a generalized Langevin 
equation:
\begin{eqnarray*}
\stackrel{..}{\phi }(Q,t)+\omega _0^2(Q)\phi (Q,t)+\mbox{$\int_0^t$}M(Q,t-t^{\prime })%
\stackrel{.}{\phi }(Q,t^{\prime })dt^{\prime }=0  && \label{langevin}
\end{eqnarray*} 
(here $\omega _0^2(Q)=[k_BT/mS(Q)]Q^2$) obtained adopting a standard 
memory function approach for the density correlator 
$\phi (Q,t)$=$<$$\rho_Q(t)\rho_{-Q}(0)$$>$$ /$$<$$\left| \rho_Q(t)\right|
^2$$>$ 
where $\rho(t)$$=$$\left( 1/N\right) \sum_ie^{-iQr_i(t)}$,
and $<$$\left| \rho_Q(t)\right| ^2$$>$=$S(Q)$, the static
structure factor.
After solving this equation, the dynamic structure factor reads \cite{bal,boo} as:

\begin{equation}
\frac{S(Q,\omega )}{S(Q)}=\frac{\pi^{-1}\omega _0^2(Q)M^{\prime }(Q,\omega )}{%
\left[ \omega ^2-\omega _0^2+\omega M^{\prime \prime }(Q,\omega )\right]
^2+\left[ \omega M^{\prime }(Q,\omega )\right] ^2}  \label{sqw}
\end{equation}

In Eq.~(\ref{sqw}), all the details of the microscopic interactions are
embodied in the (complex) ''memory function'' $M(Q,t)$, which is conveniently written
as a sum of a longitudinal ''viscous'' contribution $M_L(Q,t)$ and of a term
$M_{th}(Q,t)$ arising from the coupling to thermal fluctuations. In the $%
Q\rightarrow 0$ hydrodynamic regime, the memory function $%
M(Q,t)$$=$$M_L(Q,t)$$+$$M_{th}(Q,t)$ can be written as

\begin{eqnarray}
M(Q,t) &=&M_{L}(Q,t)+M_{th}(Q,t)  \label{memoryhyd} \\
&=&2(\eta _{L}/nm)Q^{2}\delta (t)+\left( \gamma -1\right) \omega
_{0}^{2}(Q)e^{-D_{T}Q^{2}t}  \nonumber
\end{eqnarray}
where $\eta_L$ is the longitudinal viscosity, $n$ the number density, $%
\gamma $ the specific heat ratio, and $D_{T}$ the thermal diffusivity. In
the case of liquid metals, the inaccuracy of the expression (\ref{memoryhyd}%
) at finite wavevectors is particularly evident: due to the large thermal
diffusivity of these systems, the second term in Eq.~(\ref{memoryhyd}) is in
fact instantaneous, and the lineshape deduced from Eq.~(\ref{sqw})
is the so called ''damped oscillator model'' (DHO).
This scheme is unable to account for
the quasi-elastic peaks evident from the spectra in Fig.~1, and thus must be
discarded (for further details, see Ref.\cite{scop}). A reasonable way to
improve the simple hydrodynamic, DHO-like, model (\ref{memoryhyd}) is to
allow a non-instantaneous decay of the viscous term $M_L(Q,t)$. On a
practical basis, the simplest way of doing this is to assume multiple 
exponential decay laws for the time dependence of $M_L(Q,t)$:

\begin{equation}
M_L(Q,t)=\Sigma_{i=1}^N\Delta _i^2(Q)e^{-t/\tau _i(Q)}  \label{somme}
\end{equation}
and to perform a refined analysis of the data by increasing $N$ 
until optimum agreement is obtained.
In a first step, we have
taken $N$$=$$1$, which corresponds to the usual viscoelastic model. Then we tested
the case $N$$=$$2$, which mimics the two-timescales theoretical predictions, an ansatz
for $M_L(Q,t)$ that was proposed to account for the details of $S(Q,\omega)$
in the MD simulations of Lennard-Jones liquids \cite{lev}. Despite its
arbitrary character, the exponential decay law in Eq.~(\ref{somme}) has
the advantage of yielding analytical Fourier transforms, thereby simplifying
considerably the fitting procedure.

The scattered intensity is proportional to the convolution between the
experimental resolution function $R(\omega )$ and a ''quantum-mechanical''
dynamic structure factor $S_q(Q,\omega )$, affected by the detailed balance
condition:

\begin{equation}
S_{q}(Q,\omega )={\beta \hbar \omega }/(1-e^{-\beta \hbar \omega })
S(Q,\omega )  \label{kubo}
\end{equation}
The difference between the resulting function and the experimental data 
has been minimized using a Levenberg-Marquardt procedure. $\omega _0^2(Q)$ 
has been obtained from its definition, while the quantities $\gamma $ 
and $D_{T}$ have been fixed to their experimental $Q\rightarrow 0$
values, the latter being taken from Ref. \cite{ohs}.
Consequently, the only free parameters in the fitting iterations were the
decay times, $\tau _i(Q)$, and the strengths of the viscous relaxations 
$\Delta _i^2(Q)$.
As a result, we have been able to fit our data all over the explored $Q$ range,
and to find striking differences between the use of one or two relaxation
mechanisms. The results (resolution convoluted) are shown in Fig.~1 
for the data taken at the three reported $Q$ values.
The dashed/dotted and solid lines correspond respectively to the fit with one or two
exponential functions. The fitting results provide definite evidences that the one
decay-time viscoelastic model is untenable. On the contrary, a two timescale relaxation
dynamics for the decay of $M_L(Q,t)$ successfully accounts for the observed
spectral shapes yielding  an excellent fit (full line).
The single-time viscoelastic
fits ($N$$=$$1$ in Eq.~(\ref{somme})), both initialized with free parameters
corresponding to the best values obtained in the two-times model,
give results that can be read as a sort of ''average'' of the two
processes, emphasizing in turn either the fast or the slow one.
From this comparison one notices that the faster relaxation process accounts
almost entirely for the width of the inelastic peak and for the broader part
of the quasi-elastic region, while the slower relaxation is responsible for
the narrow portion of the central peak.

\begin{figure}[f]
\centering
\vspace{-.4cm}
\includegraphics[width=.6\textwidth]{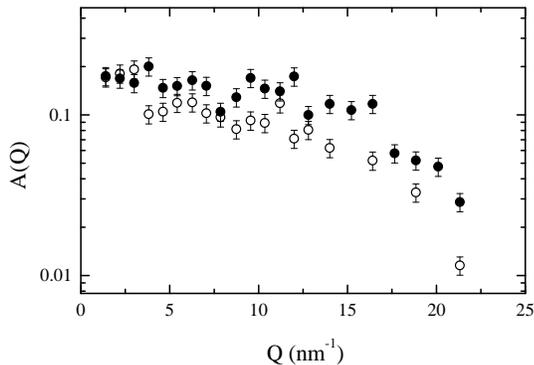}
\vspace{-2.5cm}
\hspace{5.0cm}
\caption{
Ratio, $A(Q)$, of the slow relaxation mechanism strength to the total strength. Full and open dots
refer to $T=475$ K and 600 K, respectively.}
\label{Fig2}
\end{figure}

Following the theoretical analysis of Ref. \cite{sjogr}, we shall refer to
the faster mechanism as ''microscopic'' (label $\mu $) and to the slower one as ''structural'' (label $\alpha $).
In Fig.~2 we report for the investigated temperatures
the ratio $A(Q)$ between the strengths
of the slow process $\Delta _\alpha ^2(Q)$ and the total strength $\Delta
_\alpha ^2(Q)+\Delta _\mu ^2(Q)$ as found from the fitting procedure. The
weight of the structural mechanism is seen to be largest at the smallest
wavevectors ($\sim 20\%$), and to decrease with $Q$.
This is in agreement with the expectation that the slower relaxation
process affects less the dynamics of the systems as one increases the frequency
to values sensibly larger than $1/\tau_\alpha$.

The wavevector-dependence of the relaxation times is illustrated in Fig.~3. The
microscopic time turns out to be nearly temperature independent, with values
of $\tau _\mu (Q)$ decreasing in the explored $Q$ range from $\sim 0.05$ ps
down to $\sim 0.01$ ps. The values of the relaxation time $\tau _\alpha (Q)$
associated with the structural process are instead more scattered: $\tau
_\alpha (Q)$ is always of the same order as the inverse of the energy
resolution ($\sim 0.43$ ps in the case of the (9 9 9) reflection), making
rather difficult a precise determination.

\begin{figure}[ftb]
\vspace{-.40cm}
\centering
\includegraphics[width=.4\textwidth]{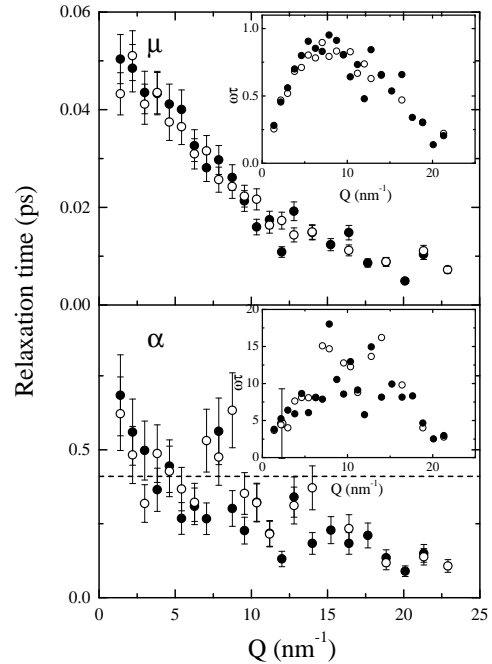}
\caption{
The times associated with fast (above) and
slow (below) relaxation as deduced from the fits. Full and open symbols
correspond to $T=475$ K and $T=600$ K, respectively. A precise determination
of the slow structural decay time is affected by the timescale corresponding
to the experimental resolution (here 3.3 meV; dashed line)}
\label{Fig3}
\end{figure}

Finally, in Fig.~4 we report the $Q$
dependence of the sound speed $c_l(Q)$$=$$\omega _l(Q)/Q$ where $\omega _l(Q)$
is the (resolution-deconvoluted) peak position of the longitudinal current
spectrum $\omega ^2S(Q,\omega )$, as obtained by the best-fitted values of
the parameters. In the same figure we also report the $Q$-dependent
isothermal sound speed $c_0(Q)$$=$$\omega _0(Q)/Q$ (expected to be the limiting
value of $c_l(Q)$ at the lowest $Q$ accessible to IXS \cite{scop} ), as well
as the ''infinite-frequency'' velocity $c_\infty (Q)$$=$$[{\gamma \omega
_0^2(Q)+\Delta _\alpha ^2(Q)+\Delta _\mu ^2(Q)}]^{1/2}/Q$ as obtained from the fit.
In principle, the quantity $c_\infty (Q)$ corresponds to an instantaneous,
solid-like, response of the liquid and can be evaluated from an integral
involving the effective interparticle potential and the pair distribution
function \cite{bal,scop}. To better enlight the different relaxation steps,
Fig.~4 also reports the values of $c_{\infty -\alpha }(Q)$$=$$[{\omega
_0^2(Q)+\Delta _\alpha ^2(Q)}]^{1/2}/Q$, the ''unrelaxed'' sound speed
associated with the presence of the only slow structural mechanism. At the
lowest wavevectors probed by IXS, the effective sound velocity $c_l(Q)$ has
already reached $c_{\infty -\alpha }(Q)$; owing to the magnitude of $A(Q)$, 
its value is, however, only about one half of the full instantaneous
response $c_\infty (Q)$. At larger values of $Q$ it is apparent a ''positive
dispersion'' of $c_l(Q)$, which reaches its maximum at $Q$$=$6 nm$^{-1}$.

\begin{figure}[ftb]
\centering
\includegraphics[width=.6\textwidth]{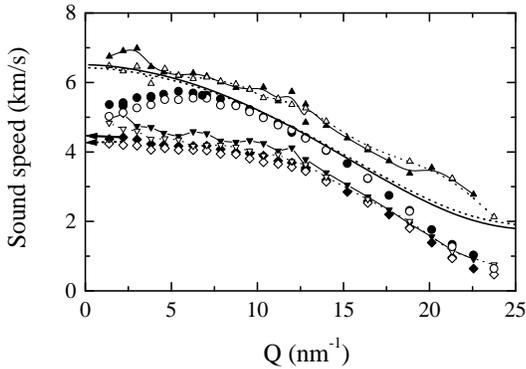}
\vspace{-2.8cm}
\caption{
Effective sound speed $c_l(Q)$ ($\bullet $)
from the peaks of the deconvoluted longitudinal current spectra as deduced
from the fits. Also reported are: the ordinary hydrodynamic sound speed [17]
($\leftarrow$ arrow), the isothermal speed $c_0(Q)$
($\blacklozenge $) and the instantaneous speed $c_\infty^{th}(Q)$
(------) as evaluated theoretically in Ref. [12]. Full and open symbols refer to $T$$=$$475$ K and 600 K, respectively.
From the fitting parameters one may also deduce: (i)
$\sqrt{\omega _0^2(Q)+\Delta _\alpha^2(Q)}/Q$ ($-\blacktriangledown -$),
the high-frequency velocity associated with
the slow process, and (ii) the total high-frequency velocity
$c_\infty (Q)$ ($-\blacktriangle -$).
The non coincidence of the latter with the theoretical
$c_\infty (Q)$ can be ascribed to the oversimplified form of the memory
function model.}
\label{Fig4}
\end{figure}

At even larger wavevectors ($Q$$\sim$$ 10$ nm$^{-1}$), the insets of Fig.~3 show
that the condition $\omega_l(Q)\tau _{\alpha ,\mu }(Q)$$\geq$$ 1$ is attained:
in this case the effective velocity $c_l(Q)$ is expected to approach the
instantaneous value $c_\infty (Q)$. Fig.~4 shows that this is indeed the case
if $c_\infty (Q)$ is evaluated theoretically from the afore-mentioned
integral. On the other hand, the fact that the values of $c_\infty (Q)$, as
calculated from the fit parameters, are larger than the ''exact'' ones is a
consequence of the oversimplified form of the two-exponential decay model in
the short-time range relevant for this comparison \cite{scop}.

Summing up,
by IXS it has been possible to detect all the main features of collective
dynamics in a simple liquid metal as lithium. Specifically, our data show the
presence of three distinct decay channels of the collective memory function.
The first one is the ''thermal'' mechanism: as in other molten alkali metals
its relevance is rather small, although not entirely negligible. Much more
important is the unambiguous evidence of two well separated timescales in
the decay of $M_L(Q,t)$. This provides a convincing experimental 
demonstration for the simultaneous presence of both fast and slow
relaxation processes in the collective dynamics of simple liquids.


\begin{references}
\bibitem{multnote} An upper estimate of the ratio between the integrated intensity 
of the one and the two phonon processes (for further details see for example \cite{monaco}) 
gives 0.05 at $Q \rightarrow 0$. Increasing the wavevector, such contribution further decreases: at 
$Q=5$nm$^{-1}$ it reduces to 0.005.
\end{references}
\end{document}